\documentclass[12pt]{article}

\usepackage{epsfig,bm,titlesec,hyperref}

\begin{document}

\title{Zero Modes in Electromagnetic Form Factors of the Nucleon in a
Light-Cone Diquark Model}

\author{\begin{tabular}{c} \normalsize Jun He,$^{1,2,3}$, Yu-bing Dong $^{1,2}$ \\
\footnotesize \sl $^1$CCAST (World Lab.), P. O. Box 8730, Beijing 100080 \\
\footnotesize \sl $^2$Institute of High Energy Physics, Chinese
Academy of Sciences, P. O. Box 918-4, 100049 Beijing, P.R. China
\\\footnotesize \sl$^3$Graduate School of the Chinese Academy of Sciences, 100049,
Beijing, P.R. China
\end{tabular}}

\date{\today}

\maketitle

\begin{abstract}
   We use a diquark model of the nucleon to calculate the
electromagnetic form factors of the nucleon described as a scalar
and axialvector diquark bound state. We provide an analysis of the
zero-mode contribution in the diquark model. We find there are
zero-mode contributions to the form factors arising from the
instantaneous part of the quark propagator, which cannot be
neglected compared with the valence contribution but can be removed
by the choice of wave function. We also find that the charge and
magnetic radii and magnetic moment of the proton can be reproduced,
while the magnetic moment of the neutron is too small. The dipole
shape of the form factors, $G^p_M(Q^2)/\mu_p$ and
$G^n_M(Q^2)/\mu_n,$ can be reproduced. The ratio $\mu G^p_E/G^p_M$
decreases with $Q^2,$ but too fast.
\end{abstract}

\section{Introduction}

Methods using relativistic formalisms directly without Foldy-Wouthuysen
reduction have been suggested~\cite{Dirac,Karmanov,Lev,Keister} to include
relativistic effects to all orders, while avoiding truncated power
expansions in $v/c$ or $p/m$. Among these relativistic formalisms, light-cone
quark models are widely used. There are several works where the electroweak
properties of mesons have been studied using light-cone quark
models~\cite{Dziembowski,Chung88,Jaus,Melo,cardarelli,Krutov}. Applications of
the Bakamjian-Thomas construction of the relativistic few-body
system~\cite{Lev,Keister,Capstick95,Chung} to extend nonrelativistic wave
functions to the relativistic domain are close to the naive quark model. Weber
$et~al.$~\cite{Konen,Stanley} give relativistic spin-flavor wave functions of
a Lorentz covariant quark model. Araujo $et.\ al.$~\cite{Araujo} connect these
wave functions with the triangle Feyman diagram and give an effective
Lagrangian of the nucleon, which corresponds to a field-theoretical
formulation of the form factors.

From the Eqs. (9,10) in Ref.~\cite{Konen} (KW, for short. See also
Eq.~(3) below.), we see that the nucleon wave function is split into
two parts as $(qq)q$. At this point, the diquark model can be
introduced naturally. The diquark model has been widely used to
study the property of baryons. For example, nucleon properties are
studied through the Bethe-Salpeter equation with scalar and
axialvector diquark correlations ~\cite{Oettel}. Kroll $et.\ al.$
study the electromagnetic form factors of the proton in the
time-like region and two-photon annihilation into proton-antiproton
in the infinite momentum frame~\cite{Kroll}.

In all investigations of baryon properties~\cite{Lev,Keister,Capstick95,Chung,
Konen,Stanley,Araujo}, zero modes have not been considered. The zero-mode
contributions can be interpreted as residues of virtual pair creation processes
 in the $q^+(=q^0+q^3)\rightarrow0$ limit of the virtual photon momentum
~\cite{Melo98,Melo99}. In the absence of zero-mode contributions,
the hadron form factor can be obtained in a straightforward way by
taking into account only the valence quark contributions. In an
effort to clarify this issue, Jaus~\cite{jausZM} and Choi $et.\
al.$~\cite{Choi} independently investigated the electroweak form
factors of spin-1 mesons in the past few years. Choi $et.\ al.$ find
that the zero-mode contributions in principle depend on the form of
the vector-meson vertex $\Gamma^\mu=\gamma^\mu-(P_v-2k)^\mu/D$; the
form factor $f(q^2)$ is found to be free from a zero-mode
contribution if the denominator $D$ contains a term proportional to
the light-front energy $(k^-)^n$ with a power $n>0$. Analogs of
$\Gamma^{\mu}$ occur for the spinor parts of the nucleon current
matrix element, as we shall discuss below.

In this paper we calculate the electromagnetic form factors of the
nucleon, taken as a scalar and axialvector diquark bound state, and
investigate the zero-mode contributions in the quark-diquark model.

In the following section, the nucleon wave function in the diquark
model and an effective quark-diquark-nucleon effective Lagrangian
are given. Then the extraction of the nucleon form factors from the
current matrix elements are given in Section~3. The discussion of
the zero mode is given in Section~4. The numerical results for the
form factors in the frame where the virtual photon momentum $q^+=0$
are given in Section~5. The summary and discussion are given in the
last section.

\section{Wave Function of the Nucleon}

In the static case the nucleon wave function with $SU(6)$ symmetry can be
written as
\begin{eqnarray}\label{Eq: SU6wf}
\frac{1}{\sqrt{2}}(\chi_\rho\Phi_\rho+\chi_\lambda\Phi_\lambda)\psi_0,
\end{eqnarray}
where $\chi_\rho\Phi_\rho$ corresponds to the scalar-isoscalar component of
the nucleon and $\chi_\lambda\Phi_\lambda$ to the vector-isovector quark pair
part. This wave function can be rewritten in quark-diquark form as
\begin{eqnarray}\label{Eq. diquarkwf}
&&|p^\uparrow\rangle=\frac{1}{\sqrt{18}}[(2V^\uparrow_{11}d^\downarrow
-\sqrt{2}V^\rightarrow_{11}d^\uparrow-\sqrt{2}V^\uparrow_{10}u^\downarrow
+V^\rightarrow_{10}u^\uparrow)\sin\alpha+3S_{00}u^\uparrow\cos\alpha],
\end{eqnarray}
where the $V$ are axial-vector diquark amplitudes and $S$ is the
scalar diquark amplitude. The explicit form of the $V$ and $S$ is
given in Ref.~\cite{Pavkovie}. If we set $\alpha=\pi/4,$  then
Eq.~(\ref{Eq. diquarkwf}) reduces to Eq.~(\ref{Eq: SU6wf}), that is,
the permutation symmetry is restored.

Following KW~\cite{Konen}, the two parts of the wave function in the
nonstatic case become
\begin{eqnarray}
&&\chi_\rho\Phi_\rho\rightarrow N_S\left(\bar{q}_1\gamma_5 \imath
\tau_2 C\bar{q}_2^T\right)\left(\bar{q}_3N_\lambda
\right);\nonumber\\
&&\chi_\lambda\Phi_\lambda\rightarrow N_V\left(\bar{q}_1\gamma^\mu
{\bm \tau}\imath \tau_2 C\bar{q}_2^T\right)\cdot \left(\bar{q}_3
\gamma_\mu {\bm \tau}\gamma_5 N_\lambda\right),
\end{eqnarray}
where a sum over quark permutations is implied. Here
$C=\imath\gamma^2\gamma^0$ is the charge conjugation matrix,
 $N_S, N_V$ are the normalization constants of the diquark wave functions.
Also $N_\lambda= [P]u_\lambda$ with $u_{\lambda}(P)$ the
Dirac-spinor of the nucleon and $[P]$ denotes $\gamma\cdot P+m_N$
with $m_N$ the nucleon mass, $q_i=[P]u_i$ with $u_i$ the quark
spinors; a factor $[P]$ with each quark spinor appears with the
Melosh rotation of the nonrelativistic spin wave function to the
light cone. This is derived in detail in Sect. IV.A of Ref.
\cite{BKW}.

\begin{figure}[h]
 \begin{center}
 \includegraphics[bb=165 585 530 750,scale=.8]{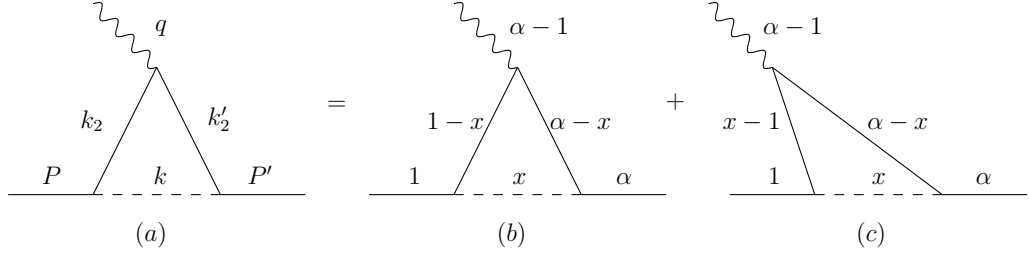}\end{center}
 \caption{Feynman diagrams for the nucleon current in case a photon
strikes the quark. The case of a photon striking the diquark is
analogous.} \label{Feynmandiagrams}
\end{figure}

Here we recognize $\bar{q}_1\gamma_5 \imath \tau_2 C\bar{q}_2^T$ and
$\bar{q}_1\gamma^\mu {\bm \tau}\imath \tau_2 C\bar{q}_2^T$ as the
scalar and axialvector diquarks from the NJL model~\cite{Reinhardt}.
Thus, the nucleon wave function can be written as (see the initial
quark-diquark vertex in Fig.~\ref{Feynmandiagrams})
\begin{eqnarray}
&&|N\rangle=f_S\varphi_S(x,k_\perp)\overline{u}(k){\cal
O}_Su_{\lambda}(P)+
f_V\varphi_V(x,k_\perp)\epsilon^i_{\alpha}\overline{u}(k) {\cal
O}^{\alpha}_{V} \frac{\tau^i}{\sqrt{3}}u_{\lambda}(P),\ \label{Eq: wfval}\\
\label{Eq: wfvaldef} &&{\cal O}_S=1,\ \ {\cal
O}_{V}^{\alpha}=(\gamma^\alpha+P^\alpha/M)\gamma_5 ,
\end{eqnarray}
where $\varphi_S,$ and $\varphi_V$ are the scalar and axialvector
radial diquark wave functions and $\epsilon_{\alpha}$ is the
axialvector diquark polarization vector, and $k^{\mu}$ is the quark
four-momentum. If coupling constants $f_S$ and $f_V$ is free
completely, there will be a common factor, which will give a non-one
charge of proton, so they must be constrained to give correct charge
of proton. Our wave function is similar to that of Kroll $et\
al.$~\cite{Kroll} and its basic diquark properties are consistent
with those of other authors~\cite{Reinhardt,Keiner}.

From Eq.~(\ref{Eq: wfval}) we can introduce an effective Lagrangian
for the quark-diquark-nucleon coupling as in Ref.~\cite{Araujo} for
the three-quark model
\begin{eqnarray}\label{Eq: effective Lagrangian}
&&{\cal L}_{qDN}=f_S\overline{\psi}(k_1){\cal
O}_S\psi_{N}(P)\phi_S(k_2)\Lambda_S+ f_V\overline{\psi }(k_1){\cal
O}^{\alpha }_V(k_2)\frac{{\bm \tau}}
{\sqrt{3}}\psi_{N}(P)\cdot{\bm \phi}_{V \alpha}(k_2)\Lambda_V,\ \\
&&{\cal O}_S=1,\ \ {\cal
O}_V^{\alpha}=(\gamma^\alpha+\imath\partial^\alpha/M) \gamma_5
,\label{Eq: Definiton of O}
\end{eqnarray}
where $\psi$, $\psi_{N}$, $\phi_S$ and ${\bm \phi}^{\alpha}_V$ are
the quark, nucleon and scalar and axialvector diquark fields,
respectively. The vertex functions $\Lambda_S$ and $\Lambda_V$ in
Eq.~(\ref{Eq: effective Lagrangian}) will be connected with the wave
functions $\varphi_S(x,k_\perp)$ and $\varphi_V(x,k_\perp)$ later.
In this work we adopt the same form for $\Lambda_S$ and $\Lambda_V.$

The Feynman rules of the diquark are presented in Table~\ref{Tab:
Feynman rule}.

\begin{table}[h]
\caption{Feynman rules of the diquark. }
\begin{center}
\begin{tabular}{c|c|c|c|c|c}
\hline \hline  $S$ & $V$ &  $SqN$   & $VqN$ & $S\gamma S$& $V\gamma
V$    \\\hline
 $\frac{\imath}{p^2-m_S^2}$ & $-\frac{\imath\delta_{ij}g_{\alpha\beta}}
{p^2-m_V^2}$ & $\imath f_S$ & $f_V{\cal
O}_V^\alpha\frac{\tau^i}{\sqrt{3}}$& $\frac{1}{3}
\Gamma_{ S}^\mu$ & $Tr[\tau_ie_q\tau_j]\Gamma_{V\alpha\beta }^\mu$ \\
\hline \hline
\end{tabular}\end{center}
\label{Tab: Feynman rule}
\end{table}

The coupling of the virtual photon to a diquark is given as~\cite{Kroll,Keiner}
\begin{eqnarray}\label{Eq: GammaS}
\Gamma_{ S}^\mu&=&-\imath(p_1+p_2)^\mu, \\
\label{Eq: GammaV}\Gamma_{V\alpha\beta}^\mu&=&\imath\left\{g_{\alpha\beta}
(p_1+p_2)^\mu-(1+\kappa)g_{\alpha}^\mu(p_1-p_2)_\beta
-(1+\kappa)g_{\beta}^\mu(p_2-p_1)_\alpha\right\}
\end{eqnarray}
with $\kappa$ the anomalous contribution of the magnetic moment of the
axialvector diquark, $p_2$ the final momentum and $p_1$ the
initial momentum of the diquark. Here, we have used the 't Hooft-Feynman
gauge parameter $\xi=1,$ as in Keiner~\cite{Keiner}, which
ensures the zero charge of the neutron.

\section{Form Factors of the Nucleon}

The electromagnetic form factors of the nucleon are conventionally
defined in terms of the electromagnetic current matrix elements,
\begin{eqnarray}
&&\langle N\lambda'|J^\mu|N\lambda\rangle=-\imath
e\bar{u}_{\lambda'}(P')\left[\gamma^\mu
F_{1}(q^2)+\frac{1}{4m_N}q_\nu[\gamma^\nu,\gamma^\mu]F_2(q^2)\right]
u_{\lambda}(P)
\end{eqnarray}
with $q=P'-P$, $P'$ and $P$ the final and initial total momentum.

The form factors $F_i(q^2)$ can be determined from the $J^+$ matrix elements,
\begin{eqnarray}
&&F_1
=\frac{m_N}{\sqrt{p^+_ip^+_f}}J^+_{\uparrow\uparrow}+\frac{m_N}
{\sqrt{p^+_ip^+_f}}\frac{m_Nq^{+2}}{Q^3_L}J^+_{\uparrow\downarrow},\nonumber\\
&&F_2=\frac{4m_N^2\sqrt{p^+_ip^+_f}}{Q^3_L}J^+_{\uparrow\downarrow},
\end{eqnarray}
where $Q^3_L=(p_{Li}p^+_f+p_{Lf}p^+_i)q^+-2p^+_fp^+_iq_L$. If
$q^+=0$, the form factor is obtained in the so-called Drell-Yan-West frame.

The Sachs form factors are
\begin{eqnarray}
&&G_E(q^2)=F_1(q^2)+\frac{q^2}{4m_N^2}F_2(q^2),\nonumber\\
&&G_M(q^2)=F_1(q^2)+F_2(q^2).
\end{eqnarray}

The form factor problem is now reduced to calculating the matrix elements
\begin{eqnarray}\label{Eq: J}
J^+_{\lambda\lambda'}=\int\frac{d^4k}{(2\pi)^4}\frac{\Lambda(k'_2)
S^+_{\lambda\lambda'}\Lambda(k_2)}{D(k'_2)D(k)D(k_2)},
\end{eqnarray}
as shown in Fig.~\ref{Feynmandiagrams}. Here $\Lambda(k'_2)$ and
$\Lambda(k_2)$ are the vertex functions $\Lambda_S$ or $\Lambda_V$ given
in Eq.~(\ref{Eq: effective Lagrangian}), and
\begin{eqnarray}
D(k) &=& k^2-m_1^2+i\varepsilon,\nonumber\\
D(k'_2) &=& k^{'2}_2-m_2^2+i\varepsilon=(P'-k)^2-m_2^2+i\varepsilon,\nonumber\\
D(k_2) &=& k^2_2-m_2^2+i\varepsilon=(P-k)^2-m_2^2+i\varepsilon.
\end{eqnarray}

The spinor part $S^+_{\lambda\lambda'}$ can be obtained from the
Feynman rules in Table~\ref{Tab: Feynman rule} and Fig.~\ref{Feynmandiagrams}.
If a photon strikes the quark, the quark is labeled as particle 2, with the
spinor parts
\begin{eqnarray}\label{Eq: SqS}
S_{qS}^\mu&=&-\imath\langle f|[k'_2]e_q\gamma^\mu[k_2]|i\rangle,
\end{eqnarray}
\begin{eqnarray}\label{Eq: SqV}
S_{qV}^\mu&=&\imath\delta_{ij}g_{\alpha\beta}\langle f|\bar{{\cal O}}_V^\alpha
\frac{\tau^{i\dag}}{\sqrt{3}}[k'_2]e_q\gamma^\mu[k_2]
{\cal O}_V^\beta\frac{\tau^j}{\sqrt{3}}|i\rangle
\end{eqnarray}
in the numerator of Eq. (\ref{Eq: J}) for the current matrix
element, where we use the notation $[p]=\gamma\cdot p+m,$ as earlier
for $[P],$ and $\bar{{\cal O}}=\gamma^0{\cal O}^\dag\gamma^0$.

If the virtual photon strikes the diquark, the diquark is labeled as particle 2
 with the spinor parts
\begin{eqnarray}\label{Eq: SDS}
S_{DS}^\mu&=&\langle f|[k]|i\rangle\frac{1}{3}\Gamma_{S}^\mu,
\end{eqnarray}
\begin{eqnarray}\label{Eq: SDV}
 S_{DV}^\mu&=&\langle f|\bar{{\cal O}}^\alpha_V\frac{\tau^{i\dag}}{\sqrt{3}}[k]
 {\cal O}^\beta_V\frac{\tau^j}{\sqrt{3}}|i\rangle
Tr[\tau_ie_q\tau_j]\Gamma^\mu_{V\alpha\beta},
\end{eqnarray}
with $\Gamma_{S}^\mu, \Gamma^\mu_{V\alpha\beta}$ from Eqs.~(\ref{Eq:
GammaS},\ref{Eq: GammaV}). Compared with the three-quark model, only
Fig.~\ref{Feynmandiagrams}(a) and ~\ref{Feynmandiagrams}(d) of
Ref.~\cite{Araujo} remain in the diquark model because we regard the
diquark as a pointlike particle so that such details of its internal
structure are no longer relevant.

From an explicit calculation we recognize the same isospin parts as
in Kroll $et.\ al.$~\cite{Kroll}. For example, the isospin part of
Eqs.~(\ref{Eq: SqS}-\ref{Eq: SDV}) for the proton is $e_u$,
$e_u+2e_d$, $e_{ud}$, $e_{ud}+2e_{uu}$ respectively, where $e_{ud}$,
$e_{dd}, e_{uu}$ are the electric charges of the diquarks as given
in Ref.~\cite{Pavkovie}.

\section{Zero Mode Analysis}

In order to search for zero modes, we now integrate the current
matrix element over $k^-$  in Eq.~\ref{Eq: J} in the valence and
nonvalence regions corresponding to Fig.~\ref{Feynmandiagrams} (b)
and (c) In frame where $q^+>0$.

\subsection{Integration in the Valence Region}

In the valence region, characterized by $0<k^+<P^+$, there is only
one pole $k^-=k^-_{on} =(m_1^2+{\bm k}^2_\perp-\imath)/k^+$ located
in the lower half of the complex $k^-$-plane. Thus, we can integrate
over $k^-$ by Cauchy's formula which yields
\begin{eqnarray}\label{Eq: intv}
J^+_{\lambda\lambda'}&=&\int\frac{d^4k}{(2\pi)^4}\frac{h(k'_2)
S^+_{\lambda\lambda'}h(k_2)}{(k^{'2}_2-m_2^2
+\imath\epsilon)(k^{'2}_2-\Lambda^{'2}+\imath
\epsilon)(k_2^2-m_2^2+\imath
\epsilon)(k_2^2-\Lambda^2+\imath\epsilon)(k^2-m_1^2 +\imath
\epsilon)}\nonumber
 \\&=& 2\pi\imath\int_0^1 dx
\frac{dk_\perp^2}{2(2\pi)^4 xx_2^2
x^{'2}_2}\frac{h(k'_2)S^+_{\lambda\lambda'}
h(k_2)|_{k^-=k^-_{on}}}{(M^2_f-M^{'2}_0)(M^2_f-M^{'2}_{\Lambda'})(M^2_i-M^2_0)
(M^2_i-M^2_{\Lambda})}
\end{eqnarray}
where $M^2_0=\frac{{\bm k}^{2}_{\perp}+m_2^2}{x_2} +\frac{{\bm
k}_{\perp}^{2}+m_1^2}{ x}$, $M^{'2}_0=\frac{{\bm
k}^{'2}_{\perp}+m_2^2}{x'_2}+\frac{{\bm k}^{'2}_{\perp}+m_1^2}{
x'}$, $M^2_\Lambda=\frac{{\bm k}^{2}_{\perp}+\Lambda^2}{x_2}
+\frac{{\bm k}_{\perp}^{2}+m_1^2}{ x}$,
$M^{'2}_{\Lambda'}=\frac{{\bm
k}^{'2}_{\perp}+{\Lambda'}^2}{x'_2}+\frac{{\bm
k}^{'2}_{\perp}+m_1^2}{ x'}$ with $x=k^+/P^+,\ x_2=1-x,\
x'=k^+/P^{'+}=x/\alpha,\ x'_2=1-x', {\bm k}'_{\perp}={\bm
k}_{\perp}-x'{\bm q}_{\perp}.$ Here we adopt ${\bm P}_\perp=0,$ and
$M_f$ and $M_i$ are the mass of the initial and final states, $i.\
e.$ the nucleon mass. We use a Pauli-Villars regulator $\Lambda$ in
$\Lambda(k_2)=h(k_2)/(k_2^2-\Lambda^2+ \imath\epsilon)$, where
$\Lambda$ plays a role of momentum cutoff, to make the integration
in Eq.~(\ref{Eq: intv}) and Eq.~(\ref{Eq: intnv}) below
finite~\cite{Melo,Choi}. From Eq.~(\ref{Eq: intv}) we see that, if
all the poles arising from $\Lambda(k_2)$, such as more
Pauli-Villars fields, lie in the upper half of the complex
$k^-$-plane, the explicit form of $\Lambda(k_2)$ has no effect on
the result in the valence region provided the definition as
Eq.~(\ref{Eq: define varphi}) are chosen to include all factor from
$\Lambda(k_2)$ to definition.  A more explicit discussion of
Pauli-Villars fields can be found in Ref. \cite{Paston97} in detail.

We can connect the integration in the valence region obtained above to
the form of Ref.~\cite{Konen} directly by using the spin sum
\begin{eqnarray}
2m[p]= \sum_\lambda u_{\lambda}(p)\bar{u}_{\lambda}(p).
\end{eqnarray}
For example,
\begin{eqnarray}
\frac{S_{qS}^\mu}{\sqrt{x'_2x_2}}&=&-\imath\langle
f|\frac{[k'_2]}{\sqrt{x'_2}}e_q\gamma^\mu\frac{[k_2]}{\sqrt{x_2}}|i\rangle
\propto\bar{u}_fu(k'_2)\frac{\bar{u}(k'_2)}{\sqrt{k^+_2}}e_q\gamma^\mu
\frac{u(k_2)}{\sqrt{{k'}^+_2}}\bar{u}(k_2)u_i
\end{eqnarray}
obviously is a current density
$\frac{\bar{u}(k'_2)}{\sqrt{k^+_2}}e_q\gamma^\mu
\frac{u(k_2)}{\sqrt{{k'}^+_2}}$ between two vertices
$\bar{u}_fu(k'_2)$ and $\bar{u}(k_2)u_i$ in Eq.~\ref{Eq: wfval}.
Then we let
\begin{eqnarray}\label{Eq: define varphi} \frac{1}{\sqrt{2(2\pi)^3}}
\frac{h(k_2)}{(m^2_N-M^2_0)x_2(m^2_N-M_\Lambda^2)}=\varphi(x_2,\bm{k}_{2\perp}),
\end{eqnarray}
where $h(k_2)/[x_2(m^2_N-M_\Lambda^2)]$ corresponds to
$\Lambda(k_2)$ before integration. $\varphi(x_2,\bm{k}_{2\perp})$ is
just the usual radial wave function in eq.~(\ref{Eq: wfval}). So if
we use a $\Lambda(k_2)$ without pole, our definition of
$\varphi(x_2,\bm{k}_{2\perp})$ is the same as the one used by Araujo
$et.\ al.$~\cite{Araujo}. And only with such a definition do the
$x$, $x_2$ $x'_2$ that are left give an invariant phase-space volume element,
$d\Gamma=d^2k_{\perp}d^2k_{2\perp}\delta(k_{\perp}+k_{2\perp})dxdx_2/(xx_2)
\delta(x+x_2-1)$ in the Drell-Yan-West frame~\cite{Konen}.

We find for the valence region results similar to Ref.~\cite{Konen}. In the
three-quark model the results with the integration as in this work are the
same as the results in Ref.~\cite{Konen} because all terms with $[p]$ are
on-shell so that they can be evaluated as $\sum_\lambda u_{\lambda}(p)
\bar{u}_{\lambda}(p)$.

In the $q^+=0$ frame, that is, the Drell-Yan-West frame,
$J^+_{\lambda\lambda'}$ can be obtained by the limit
\begin{eqnarray}
J^{+DYW}_{\lambda\lambda'}=\lim_{\alpha\rightarrow1}J^+_{\lambda\lambda'}.
\end{eqnarray}

In the valence region corresponding to Fig.~\ref{Feynmandiagrams} (b), when
$\alpha\rightarrow1,$ then $x'\rightarrow x$, $x'_2\rightarrow x_2$, and so on.
 We then find that $J^{+DYW}_{\lambda\lambda'}$ is obtained from the wave
function in Eq.~(\ref{Eq. diquarkwf}) and the free quark current density
$\frac{\bar{u}(k'_2)}{\sqrt{k^+_2}}e_q\gamma^\mu
\frac{u(k_2)}{\sqrt{{k'}^+_2}}$ or the diquark current in
Eqs.~(\ref{Eq: GammaS}, \ref{Eq: GammaV}) as usual. From
Eq.~(\ref{Eq: intv}) we see that, if a pole of $\Lambda(k_2)$ arises
from $(k^{2}_2-m^2+\imath\epsilon)$, where $m=m_2,\ \Lambda,\ etc$,
it will lie in the upper half of the complex $k^-$-plane. Thus, the
explicit form of $\Lambda(k_2)$ has no effect on the integration in
the valence-quark region provided the diquark structure function
$h(k_2)$ is chosen appropriately, as will be discussed below.

\subsection{Integration in the Nonvalence Region}

The nonvalence region is characterized by $P^+<k^+<P^{'+}$
kinematically. We can recognize Fig.~\ref{Feynmandiagrams}~(c) as a
pair creation process in the nonvalence region from the orientation
of the momentum. Thus, the nonvalence contribution corresponds to a
Z-diagram in the three-quark model if the virtual photon strikes a
quark and corresponds to a double Z-diagram if the virtual photon
strikes the diquark. Thus dynamically, the nonvalence region can no
longer be described by a wave function. In the Drell-Yan-West frame
there is no such contribution to the $J^+$ current component, so it
must vanish when $q^+\rightarrow0$, that is, $\alpha\rightarrow1$.
Otherwise there would be a so-called zero mode contribution.

In the nonvalence region there are two poles located in the lower half of the
complex $k^-$-plane when we integrate the current density over $k^-$ by
Cauchy's formula to yield
\begin{eqnarray}\label{Eq: intnv}
J^+_{\lambda\lambda'}&=&\int
\frac{d^4k}{(2\pi)^4}\frac{h(k'_2)S^+_{\lambda\lambda'}h(k_2)}{(k^{'2}_2-m_2^2
+\imath\epsilon)(k^{'2}_2-\Lambda^{'2}+\imath
\epsilon)(k_2^2-m_2^2+\imath \epsilon)(k_2^2-\Lambda^2+\imath
\epsilon)(k^2-m_1^2+\imath \epsilon)}\nonumber \\
&=& 2\pi\imath\int_1^\alpha dx
\frac{dk_\perp^2}{2(2\pi)^4xx^{''2}x'_2}\frac{1}{\Lambda^{'2}-m_2^2}
\{\frac{h(k'_2)S^+_{\lambda\lambda'}h(k_2)|_{k^-=p_f^--k^{'-}_{2on}}}
{(M^2_f-M^{'2}_0)(q^2-M_{m_2m_2})(q^2-M_{m_2\Lambda})}\nonumber \\
&-&\frac{h(k'_2)S^+_{\lambda\lambda'}h(k_2)|_{k^-=p_f^--k^{'-}_{2\Lambda}}}
{(M^2_f-M^{'2}_{\Lambda'})(q^2-M_{\Lambda'm_2})(q^2-M_{\Lambda'\Lambda})}\},
\end{eqnarray}
where $x''=x_2/(1-\alpha),\ {\bm k}''_{\perp}={\bm k}_\perp-x''{\bm
q}_\perp$, $M_{m_2m_2}=\frac{ {\bm k}^{''2}_{\perp}+m^{2}_2}{1-x''}
+\frac{{\bm k}^{''2}_{\perp}+m_2^2}{x''}$, $M_{m_2\Lambda}
=\frac{{\bm k}^{''2}_{\perp}+m_2^2}{1-x''}+\frac{ {\bm
k}^{''2}_{\perp} +\Lambda^2}{x''}$, $M_{\Lambda'm_2}=\frac{{\bm
k}^{''2}_{\perp} +\Lambda^{'2}}{1-x''}+\frac{ {\bm
k}^{''2}_{\perp}+m_2^2}{x''}$, $M_{\Lambda'\Lambda}=\frac{{\bm
k}^{''2}_{\perp}+\Lambda^{'2}}{1-x''} +\frac{ {\bm
k}^{''2}_{\perp}+\Lambda^2}{x''}$.

If there are more Pauli- Villars fields used, that is,
$\Lambda(k_2)=h(k_2)/\prod_i(k_2^2-\Lambda_i^2+ \imath\epsilon)$, we
can use relation
\begin{eqnarray}\frac{1}{(k_2^2-\Lambda_i^2+
\imath\epsilon)(k_2^2-\Lambda_j^2+
\imath\epsilon)}=\frac{1}{\Lambda_i^2-\Lambda_j^2}\left(\frac{1}{k_2^2-\Lambda_i^2+
\imath\epsilon}-\frac{1}{k_2^2-\Lambda_j^2+ \imath\epsilon}\right)
\end{eqnarray}
to split the integration to the sum of integration as Eq.~(\ref{Eq:
intnv}).

Now we connect the integration in the nonvalence region contribution
to the form of Ref.~\cite{Konen}. In the valence region we used the
spin sum $ 2m[p]=\sum_\lambda u_{\lambda}(p)\bar{u}_{\lambda}(p)$.
However, in the nonvalence region we have $x_i=x-1$ for initial
part, so we should define $k_2$ as $k-P,$ and not $P-k$ as $k_2$,
that is, all $k_2$ should be replaced by $-k_2$. For example,
\begin{eqnarray}
\frac{S_{qS}^\mu}{\sqrt{x'_2x_2}}&=&-\imath\langle
f|\frac{[k'_2]}{\sqrt{x'_2}}e_q\gamma^\mu\frac{[-k_2]}{\sqrt{x_2}}|i\rangle
\propto\bar{u}_fu(k'_2)\frac{\bar{u}(k'_2)}{\sqrt{k^+_2}}e_q\gamma^\mu
\frac{v(k_2)}{\sqrt{{k'}^+_2}}\bar{v}(k_2)u_i
\end{eqnarray}
 using the spin sum
\begin{eqnarray}
2m[-p]= -\sum_\lambda v_{\lambda}(p)\bar{v}_{\lambda}(p).
\end{eqnarray}
Obviously, this is a current density
$\frac{\bar{u}(k'_2)}{\sqrt{k^+_2}}e_q\gamma^\mu
\frac{v(k_2)}{\sqrt{{k'}^+_2}}$ between two vertices
$\bar{u}_fu(k'_2)$ and $\bar{v}(k_2)u_i$. Hence the nonvalence
region contribution is just the virtual pair creation process.

Because $h(k_2)$ functions in both regions are from the same one in
Eq.~(\ref{Eq: J}) we can still define $h(k_2)$ as
\begin{eqnarray}\label{Eq: define varphi nv} \frac{1}{\sqrt{2(2\pi)^3}}
\frac{h(k_2)}{(m^2_N-M^2_0)x_2(m^2_N-M_\Lambda^2)}=\varphi(x_2,\bm{k}_{2\perp}).
\end{eqnarray}
However in Eq.~(\ref{Eq: intnv}) only $(M^2_f-M^{'2}_0)$ and
$(M^2_f-M^{'2}_{\Lambda'})$ occur for the final state. The vertex in
the initial state is the so-called no-wave function vertex which,
via Eq.~(\ref{Eq: Definiton of O}), has a spin part different from
the nucleon wave function but still a radial nucleon wave function.
This vertex was also studied in Refs.~\cite{cyc,crj,cwh}. In the
work by Choi et. al. they use BS equation to connect the nonwave
function vertex with the normal vertex with a kernel. In our work,
we do not make explicit calculation of nonvalence contribution, so
we use the results without using BS equation directly.

\subsection{Recognition of Zero Modes}

Now we proceed to look for a zero mode of the nucleon, that is, the
spin-1/2 case. The zero mode contribution is defined as in Ref.~\cite{Choi}
\begin{eqnarray}\label{Eq: ZMde}
J^+_{\lambda\lambda'Z.M.} &=&
\lim_{\alpha\rightarrow1}J^+_{\lambda\lambda'nv}.
\end{eqnarray}
No singularities, hence zero modes, arise at $x=0.$ To isolate the
zero mode contribution at $x=1$ or $x=\alpha$, we first make a
transformation of the longitudinal integration variable,
\begin{eqnarray}\label{Eq: intTrans}
J^+_{\lambda\lambda'Z.M.} &\sim&
\lim_{\alpha\rightarrow1}\int_1^\alpha dx
\frac{x(\alpha-x)[x''(1-x'')]^2}{xx^{''2}x'_2}
h(k'_2)S^+_{\lambda\lambda'}h(k_2)[...]\nonumber \\
&=&\lim_{\alpha\rightarrow1} \int_1^\alpha dx
\frac{\alpha(\alpha-x)^2}{(1-\alpha)^2}
h(k'_2)S^+_{\lambda\lambda'}h(k_2)[...]
\nonumber \\
&=& \lim_{\alpha\rightarrow1}\int_0^1 dz\, \alpha(\alpha-1)(1-z)^2
h(k'_2)S^+_{\lambda\lambda'}h(k_2)[...]
\end{eqnarray}
where $x=1-z(1-\alpha)$ and $x''=(1-x)/(1-\alpha)=z.$ So, both
$\alpha-x$ and $1-x$ can give a factor $1-\alpha$. Hereafter we
identify $\alpha-x$ and $1-x,$ when we consider the zero mode
contribution. Our analysis at the endpoint $1-x$
 will be based on counting powers of $1-x$ in the integrand of
Eq.~(\ref{Eq: intTrans}).

Now we take
\begin{eqnarray}\label{Eq: define varphi nv} \varphi(x_2,\bm{k}_{2\perp})=
\frac{1}{\sqrt{2(2\pi)^3}}
\frac{\Lambda^2}{(m^2_N-M^2_0)x_2(m^2_N-M_\Lambda^2)}.
\end{eqnarray}
that is, $h(k_2)=\Lambda^2$ and $h(k'_2)=\Lambda'^2$, which is just
the form used by Choi et. al~\cite{Choi} as a starting point of
their analysis, because this is the simplest choice and it means
that there is no complicated internal structure in the vertex. Then
the key to recognizing a zero-mode contribution is the spinor part
$S^+_{\lambda\lambda'}$ in Eq. (\ref{Eq: J}). Using Table IV in
Ref.~\cite{Konen}, it is easy to count $(1-x)^n$ factors in
$S^+_{\lambda\lambda'},$ with $(1-x)$ coming from $x'_2$ or $x_2.$
Every $[k_{2on}]$ or $[k'_{2on}]$ in $S^+_{\lambda\lambda'}$ leads
to a term with the factor $(1-x)^{-1}$, where $k_{2on}$ or
$k'_{2on}$ are the momenta before and after the photon strikes the
quark or diquark. Adjacent $[k_{2on}]$ and $[k'_{2on}]$ terms (that
is, without another $[p_{on}]$ between them) give a single factor
$(1-x)$ because $A^\pm_{ik}$ or $K^{R,L}_{ik}$ in Eq. (A11) in
Ref.~\cite{Stanley} will give a $(1-x)$ factor when $q\rightarrow0$.
The $\gamma^+$ and $\gamma^-$ between $[k_{2on}]$ and $[k'_{2on}]$
give a factor $(1-x)$ and $(1-x)^{-1},$ respectively, and $\gamma^+$
between $[k_{2on}]$ and $[k'_{2on}]$ gives a factor $(1-x)$. In
contrast, $\gamma^+$ and $\gamma^-$ in other cases and other
$\gamma$'s, such as $\gamma_5$, $\gamma^L$, do not give rise to
terms with factors $(1-x)^{n}$.

Applying this power counting technique to the following example,
when the spinor part is given by~\cite{Choi}
\begin{eqnarray}\label{EQ: Meson}
S^+_h&=&Tr[[k'_2]\gamma^+(1-\gamma_5)[k_2][k]\gamma_5\epsilon^*\cdot
\Gamma]\nonumber \\
&=&Tr[[k'_{2on}]\gamma^+(1-\gamma_5)[k_{2on}][k_{on}]\epsilon^*\cdot\Gamma
\gamma_5]\nonumber\\&+&Tr[[k'_{2on}]\gamma^+(1-\gamma_5)[k_{2on}]
\gamma^+\epsilon^*\cdot\Gamma\gamma_5]\frac{1}{2}(k^--k_{on}),
\end{eqnarray}
where $\Gamma^\mu=\gamma^\mu-(k+k')^\mu/D$ and $\epsilon_{\alpha}$
is a polarization vector, then in both lines after the second
equality sign, the $(1-x)^{-2}$ from $[k'_{2on}]$ and $[k_{2on}]$ is
canceled by $\gamma^+(1-\gamma_5)$ and the adjacent $[k'_{2on}]$ and
$[k_{2on}]$ terms. For the $\epsilon^*\cdot\gamma$ part of
$\epsilon^*\cdot\Gamma$ in the first line, there will be $(1+x)$ and
$(1-x)^0$ from $\epsilon^{*-}\gamma^ +\gamma_5$ (or
$\epsilon^{*-}\gamma^-\gamma_5$) and ${\bm \epsilon}^*\cdot {\bm
\gamma}\gamma_5$ respectively. However $\epsilon^*\cdot(k+k')$ in
the $D$ term will give a $(1-x)^{-1}$ term. For the
$\epsilon^*\cdot\gamma$ part of $\epsilon^*\cdot\Gamma$ in the
second line, the $\gamma^+$ from $\epsilon^*\cdot\Gamma$ disappears
because of the other $\gamma^+$ using $\gamma^{+2}=\gamma^+$. The
trace will let $[k'_{2on}]$ and $[k_{2on}]$ become adjacent again.
So, the $(1-x)^{-1}$ from $\frac{1}{2} (k^--k_{on})$ will be
canceled. Then there will be $(1-x)$ and $(1-x)^0$ from $\gamma^+
{\bm \epsilon}^*\cdot{\bm \gamma}\gamma_5$ and
$\gamma^+\epsilon^{*+}\gamma^- \gamma_5,$ respectively. Thus, a zero
mode has to come from $D$ in $\Gamma^\mu$ in $(1-x)^0$ order.

We now use the same method to analyze $S_{qS}, S_{qV}, S_{DS},
S_{DV}$ of Eqs.~(\ref{Eq: SqS}-\ref{Eq: SDV}).

For $S_{qS}$ the analog of Eq.~(\ref{EQ: Meson}) is
$[k'_2]e_q\gamma^\mu[k_2]$ from Eq.~(\ref{Eq: SqS})
 in which the $(1-x)^{-2}$ from $[k'_{2on}]$ and $[k_{2on}]$ is canceled by the
$\gamma^+$ matrix element and the adjacent $[k'_{2on}]$ and
$[k_{2on}]$ terms. The same applies to $S_{qV}$, where the analog of
Eq.~(\ref{EQ: Meson}) is $\bar{{\cal O}}_V^\alpha
[k'_2]e_q\gamma^\mu[k_2]{\cal O}_V^\alpha$ from Eq.~(\ref{Eq: SqV}).
For $S_{DS}$ in Eq.~(\ref{Eq: SDS}), the instantaneous part of the
term $[k]$, which comes from the quark propagator, is
$\gamma^+(M^2-M^{'2})/2P^{'+}$, which will give a $(1-x)^{-1}$
factor. This $(1-x)^{-1}$ factor will be canceled by that from
$(k_2+k'_2)^+$. Since there is already a factor $1-\alpha$ in
Eq.~\ref{Eq: intTrans}, there will be no zero mode from $S_{qS}$,
$S_{qV}$ and $S_{DS}$.

Now we consider the spinor part $S_{DV}$ whose analog of
Eq.~(\ref{EQ: Meson}) is $\bar{{\cal
O}}^\alpha_V\frac{\tau^{i\dag}}{\sqrt{3}}[k]{\cal O}^\beta_V
\frac{\tau^j}{\sqrt{3}}Tr[\tau_i e_q\tau_j]\Gamma^\mu_{\alpha
\beta}$ of Eq.~(\ref{Eq: SqV}) with $\Gamma^\mu_{\alpha \beta}$ from
Eq.~(\ref{Eq: GammaV}). As far as $g_{\alpha\beta}(p_1+p_2)^\mu$ in
$\Gamma^\mu_{\alpha\beta}$ is concerned, $(p_1+p_2)^\mu$ can cancel
the $(1-x)^{-1}$ factor from $[k]$ as in $S_{DS}$. However, for
$(1+\kappa)(g_{\alpha}^\mu q_\beta-g_{\beta}^\mu q_\alpha)$, the
factor $(1-x)^{-1}$ can not be canceled. Hence, if we remember that
$h(k'_2)$ and $h(k_2)$ are constant, as in Choi $et.\
al.$~\cite{Choi}, there is a zero mode contribution in $(1-x)^0$
order from $S_{DV}$.

From the above analysis, with a constant $h(k_2)$, there will be zero
mode. Obviously, if  we choose $\varphi(x_2,\bm{k}_{2\perp})$
to let $h(k_2)$ have a pole, that is, have a factor $(1-x)^{n}$ with $n<0$,
there will be a higher order zero mode while if $h(k_2)$ has a factor
$(1-x)^{n}$ with $n>0$, which could reduce $h(k_2)$ after the
transformation in Eq.~(\ref{Eq: intTrans}) to zero, the
zero modes will disappear.

In the current matrix element, Eq.~(\ref{Eq: intnv}), the term $[k]$
comes from the quark propagator whose instantaneous part is
$\gamma^+(M^2-M^{'2})/2P^{'+}$. Hence the corresponding
instantaneous spinor part
\begin{eqnarray}\label{Eq: ZMDV}
 S_{DVinst}^+&=&\langle f|\bar{{\cal
 O}}^\alpha_V\frac{\tau^{i\dag}}{\sqrt{3}}[k]_{inst}
 {\cal O}^\beta_V\frac{\tau^j}{\sqrt{3}}|i\rangle
Tr[\tau_ie_q\tau_j]\Gamma^+_{\alpha\beta}\nonumber\\
&=&\imath\langle N|\frac{\tau^{i\dag}}{\sqrt{3}}
\frac{\tau^j}{\sqrt{3}}|N\rangle
Tr[\tau_ie_q\tau_j](M^2-M^{'2})/2P^{'+}\nonumber\\
&\cdot&\bar{u}_{\lambda'}\gamma_5(\gamma^\alpha+
P^{'\alpha}/M)\gamma^+ (\gamma^\beta+ P^\beta/M)\gamma_5u_\lambda\nonumber\\
&\cdot&
\left\{g_{\alpha\beta}(p_1+p_2)^++(1+\kappa)g_{\alpha}^+(k'_2
-k_2)_\beta -(1+\kappa)g_{\beta}^+(k'_2
-k_2)_\alpha\right\}\nonumber\\
&\sim&\imath(1+\kappa)\langle N|\frac{\tau^{i\dag}}{\sqrt{3}}
\frac{\tau^j}{\sqrt{3}}|N\rangle
Tr[\tau_ie_q\tau_j](M^2-M^{'2})/2M\nonumber\\
&\cdot& \bar{u}_{\lambda'}[\gamma^+\gamma\cdot q- \gamma\cdot
q\gamma^+- \gamma^+q^2 /M]u_\lambda\nonumber\\
&=&C(1+\kappa)(M^2-M^{'2}) \left\{\begin{array}{ll}
P^+q^2/M^2,&\lambda'=1/2, \lambda=1/2\\
2P^+q_L/M,&\lambda'=1/2, \lambda=-1/2\\
\end{array}\right.
\end{eqnarray}
where $C=\imath\langle N|\frac{\tau^{i\dag}}{\sqrt{3}}
\frac{\tau^j}{\sqrt{3}}|N\rangle Tr[\tau_ie_q\tau_j]$,
$M^{'2}=M^{'2}_0$ or $M^{'2}_{\Lambda'}$. After $\sim$, we have
omitted all higher order terms of $(1-\alpha)$ because they will
vanish when $q^+\rightarrow0$. Hence there is clearly a zero mode in
$S_{DV},$ when $q^+\rightarrow0$.

\subsection{Numerical Estimate of the Zero Mode}

\begin{figure}[h]
  \includegraphics[bb=0 5 330 250,scale=.7]{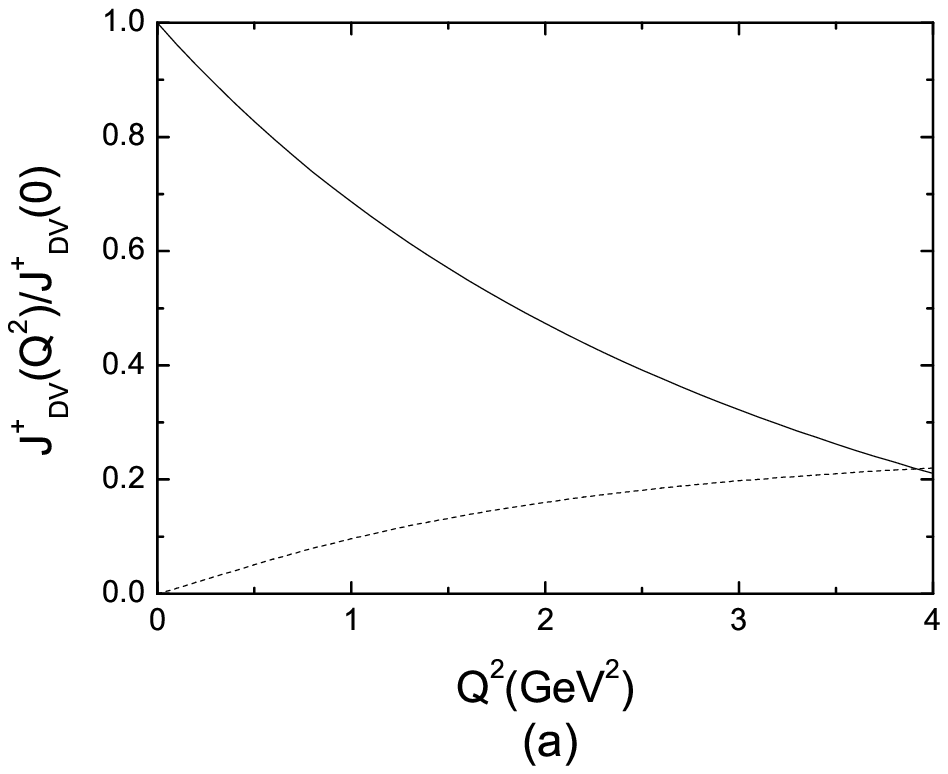}
  \includegraphics[bb=0 5 330 250,scale=.7]{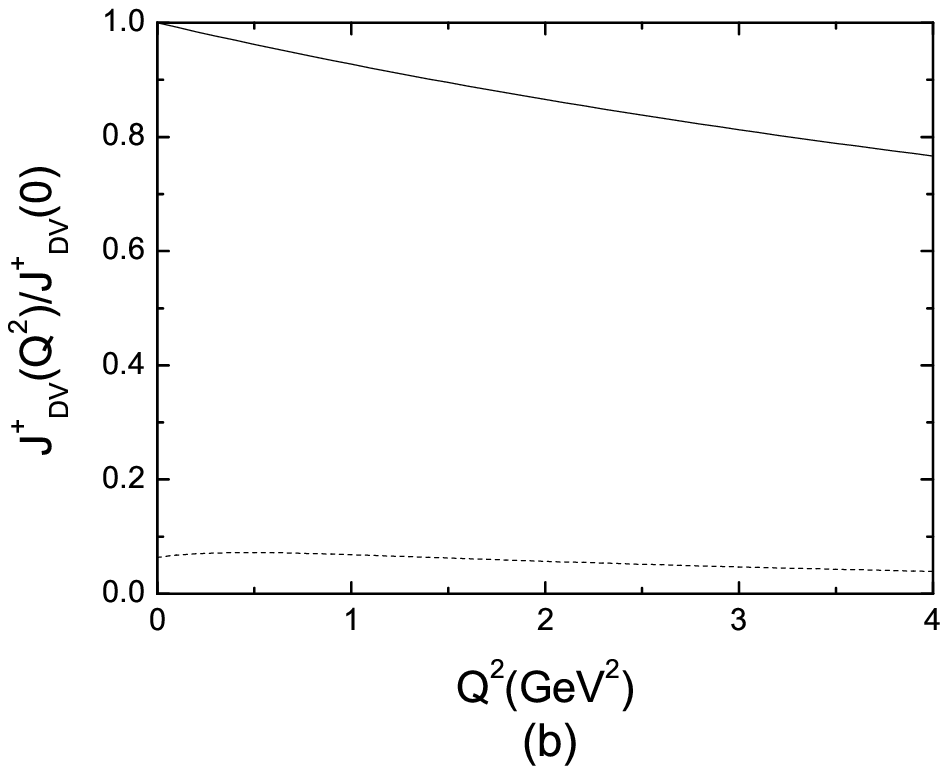}
\caption{$J^+_{DV}(Q^2)$ in the valence and nonvalence regions divided by
$J^+_{DV}(0)$ in the valence region. (a) is for the case
$\lambda=1/2$,$\lambda'=1/2$ and (b) for the case
$\lambda=1/2$,$\lambda'=-1/2$. The full line is for the valence
contribution, the dashed line for the nonvalence contribution.}
 \label{Fig: zero mode}
\end{figure}

To give an estimate of the zero mode, in Fig~\ref{Fig: zero mode} we
present the valence and nonvalence contribution of
$J^+_{DV}(Q^2)/J^+_{DV}(0)$ in the $q^+\rightarrow0$ limit with
$h(k'_2)$ and $h(k_2)$ taken as constants using Eqs.~(\ref{Eq: intv},
\ref{Eq: intnv}, \ref{Eq: ZMDV}). The valence contribution
can be calculated using Eqs.~(\ref{Eq: SDV}, \ref{Eq: intv}) directly.
The nonvalence contribution can be obtained from  Eqs.~(\ref{Eq: SDV},
\ref{Eq: intnv}) and the terms involving a zero mode have been given in
Eq.~(\ref{Eq: ZMDV}). Here we use a mass $m_q=0.43~GeV$ for the $u$ ($d$) quark
as in Refs.~\cite{Melo99,Choi} and a diquark mass of
$m_{D}=0.6~GeV$. We also adopt $\Lambda=\Lambda'=1.5~GeV$.
For both cases, the nonvalence contribution is visible. In the case
$\lambda=1/2$,$\lambda'=1/2$ in Fig. 2(a), the nonvalence
contribution increases with $Q^2$ and will cross the valence
contribution at about $-q^2=4~GeV^2$. The non valence contributions
will decrease slowly after about $Q^2=8~GeV^2$. Hence the zero mode
is far from negligible.

This power counting method we used is not required in the spin-1
meson case because the trace can be evaluated. However in the
spin-1/2 case, this is difficult to do especially in the three quark
model.

\section{Numerical Results}

In Sec. 4, we have found that if the wave functions are chosen to
let $h(k_2)$ and $h(k'_2)$ be constants, there will be a zero mode
arising when $q^+\rightarrow0$, $i.\ e.$ the Drell-Yan-West frame is
used, which indicates that the nonvalence contribution must be
considered. In the meson case, there are many works calculating a
nonvalence contribution~\cite{cyc,crj, cwh}. It would also be
interesting to study the nonvalence contribution in the timelike
region where $q^2>0$ so that we must use a frame with $q^+>0$.
However in this paper, we focus on the spacelike region where
$q^2<0$ so that Drell-Yan-West frame can be used. In case $h(k_2)$
and $h(k'_2)$ are not constant, from the analysis in Sec. 4 we can
choose these wave functions to remove zero modes.

We calculate numerical results in the Drell-Yan-West frame defined
by $q^+=0$. In this work we use for the wave functions
$\varphi_S(x,k_\perp)$ and $\varphi_V(x,k_\perp)$ in Eq.~4 both the
Gaussian parametrization
\begin{eqnarray}\label{Eq: replaceGau}
\varphi(x,k_\perp)=N\ \exp(-M^2_0/8\beta^2) \ ,
\end{eqnarray}
and the negative power parametrization indicated by the form factor
of Ref~\cite{Santopinto}.
\begin{eqnarray}\label{Eq: replaceNP}
\varphi(x,k_\perp)= N\ (\omega^2+M^2_0)^{-3.5} \ ,
\end{eqnarray}
where $N$ is a normalization constant. Inserting the Eq.~(\ref{Eq:
replaceGau})
 to Eqs.~(\ref{Eq: define varphi nv}, \ref{Eq: intnv}),
from Eq.~(\ref{Eq: intTrans}), we can find that the Gaussian form
removes any zero mode because of $M^2_0\sim (x-1)^{-1}$ (Here we
note that in the nonvalence region $x_{2i}=x-1\geq 0$ and
$x_{2f}=\alpha-x\geq 0$ as given in fig.~\ref{Feynmandiagrams}(b),
which ensures $M_0^2>0$ when $q^+\rightarrow0$). The negative power
shape wave functions will also remove the zero mode because $h(k_2)$
and $h(k'_2)$ determined by this wave function will generate a
$(x-1)^5$ factor in Eq.~(\ref{Eq: intTrans}).

The mass of the quark and diquark are chosen as $m_N/3$ and
$2m_N/3,$ respectively, and $\kappa=1.6$ as in Ref.~\cite{Keiner}
(in fact, the results are not sensitive to $\kappa$). From our
calculations we find the results are more sensitive to the coupling
constants $f_S$ and $f_V$. In the works using direct Melosh
rotation\cite{Capstick95,Cardarelli}, the proton charge can be
obtained automatically. However, as discussed in Sec 2, we let the
scalar diquark coupling constant $f_S$ determined by $G_C^p(0)=1$.
In Refs. \cite{Konen,Gross}, they also normalized the wave function
to charge one of proton. The free parameters left are $f_V$, $\beta$
for the Gaussian type and $\omega$ for negative power type wave
functions. The zero charge of the neutron is guaranteed by the
formula we used, that is, it is not fitted.

The parameters characterizing various fits are given in Table~II.
\begin{table*}[h]\label{Tab: parameters}
\caption{Parameters for Gaussian type wave function (Gau) and for
negative power one (NP).}
\begin{center}
\begin{tabular}{l|cccccc}
\hline \hline &   $m_q(MeV)$& $m_D(MeV)$ & $\kappa$ &$\beta(MeV)$ &
$\omega(MeV)$
\\\hline
  Gau        &  $313$     &  $626$       &  $1.6$    &$225$          &$--$\\
  NP         &  $313$     &  $626$       &  $1.6$    &$--$           &$100$\\
\hline \hline
\end{tabular}\end{center}

\end{table*}

We present the charge and magnetic radii and magnetic moment of the
nucleon with different $f_V$ in Table~III.

\begin{table*}[h]\label{Tab: radii and MM}
\caption{Charge and magnetic radii and magnetic moments for
different values $f_V$. The data are from
Refs~\cite{PDG,Sick,Kubon,Hammer}.}
\begin{center}
\begin{tabular}{c|c|cccccc}
\hline \hline  &$f_V (MeV)$&  $r^p_C (fm)$ & $r^p_M (fm)$ &
$r^{2n}_C (fm^2)$ &$r^n_M (fm)$ &$\mu_p(\mu_N)$ &$\mu_n(\mu_N)$
\\\hline
  Gau I  &$80$  &  $0.804$ & $0.834$ &  $-0.262$   &$0.861$ &$2.225$  &$-1.009$    \\
  NP I   &$80$  &  $0.821$ & $0.855$ &  $-0.274$   &$0.882$ &$2.228$  &$-1.013$    \\
  Gau II &$100$ &  $0.821$ & $0.820$ &  $-0.284$   &$0.850$ &$2.427$  &$-1.075$    \\
  NP II  &$100$ &  $0.836$ & $0.837$ &  $-0.295$   &$0.868$ &$2.428$  &$-1.078$    \\
  Gau III&$130$ &  $0.854$ & $0.796$ &  $-0.325$   &$0.832$ &$2.814$  &$-1.201$    \\
  NP III &$130$ &  $0.872$ & $0.816$ &  $-0.339$   &$0.852$ &$2.811$  &$-1.203$    \\\hline
  Exp.   &$--$ &   $0.870(8)$ & $0.855(35)$ &  $-0.116(2)$   &$0.873(11)$ &$2.793$  &$-1.913$\\ \hline \hline
\end{tabular}\end{center}

\end{table*}

As in the work of Kroll et al.~\cite{Kroll}, the results for the proton are
better than those for the neutron. The charge and magnetic radii and
magnetic moment are close to the data with $f_V=130~MeV$. The
magnetic moment of the neutron is too small, while the charge radius
is too large, as is the case with other authors.

\begin{figure}[h]
  \includegraphics[scale=.75]{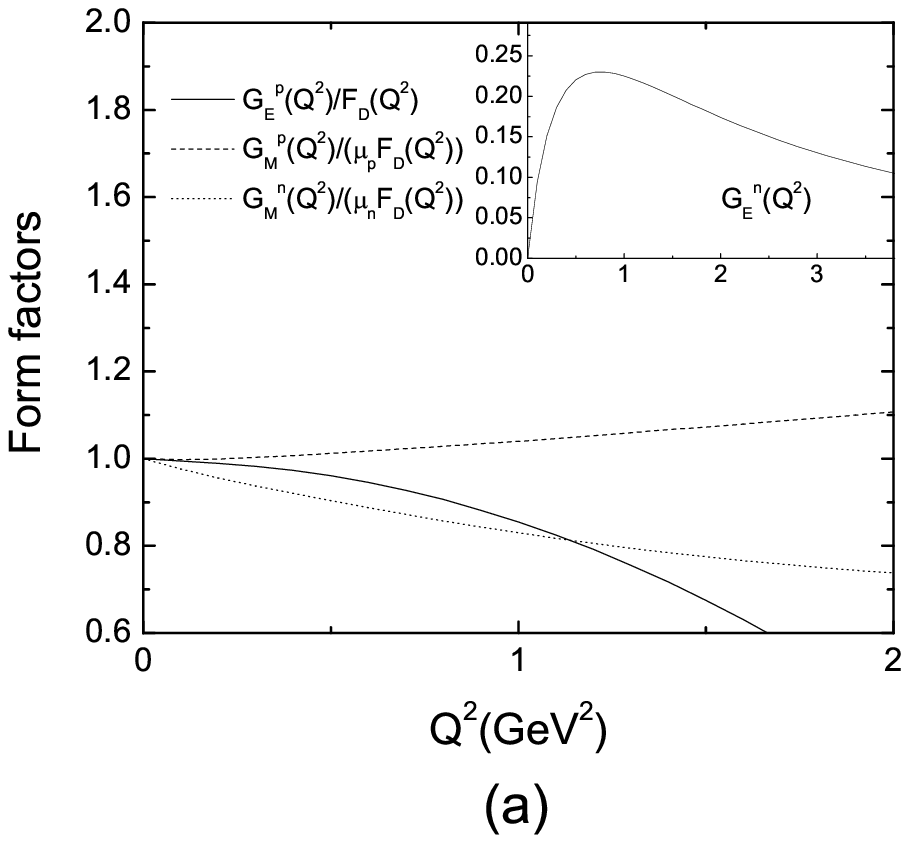}
  \includegraphics[scale=.75]{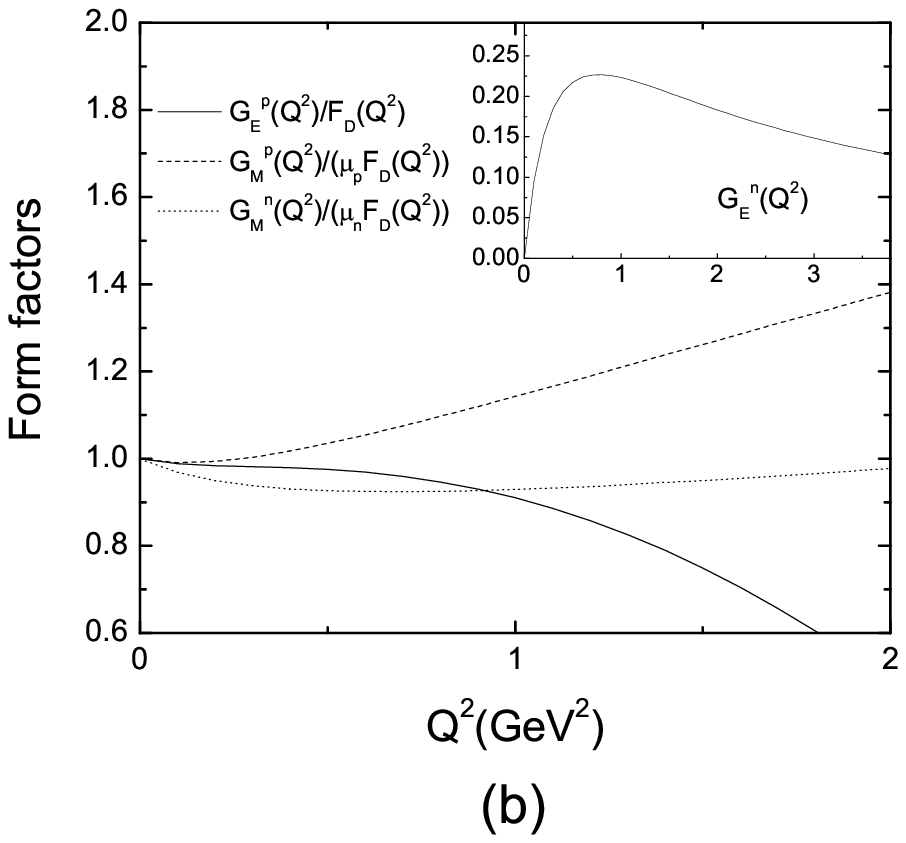}
\caption{Form factors of the nucleon. (a) and (b) are for the
Gaussian type and negative power type wave functions with
$f_V=100~MeV,$ respectively.}\label{Fig:Formfactors}
\end{figure}

The electromagnetic form factors of the nucleon are shown in
Fig.~\ref{Fig:Formfactors}. Since the experimental magnetic form factors and
the electric form factor of the proton are well described by a dipole fit
\begin{eqnarray}\label{Eq: dipole}
G^p_E(Q^2)=G^p_M(Q^2)/\mu_p=G^n_M(Q^2)/\mu_n=F_D(Q^2)=\left(1+\frac{Q^2}
{m_D^2}\right)^{-2}
\end{eqnarray}
with $m_D^2=0.71~GeV^2$.

In Fig.~\ref{Fig:Formfactors}, we present $G^p_E(Q^2)$,
$G^p_M(Q^2)/\mu_p$ and $G^n_M(Q^2)/\mu_n$ divided by $F_D(Q^2)$ and the
charge form factor of the neutron, $G^n_E(Q^2)/\mu_p$, with
$f_V=100~MeV$. The magnetic form factors of the proton and neutron are
close to the dipole form with both wave functions while
$G^p_M(Q^2)/\mu_n$ is lower than the dipole form, which is consistent
with Jefferson Lab data~\cite{Jones,Gayou01,Gayou02}. The magnetic
form factor of the neutron is a bit too high.

\begin{figure}[h]
  \begin{center}
  \includegraphics[scale=.8]{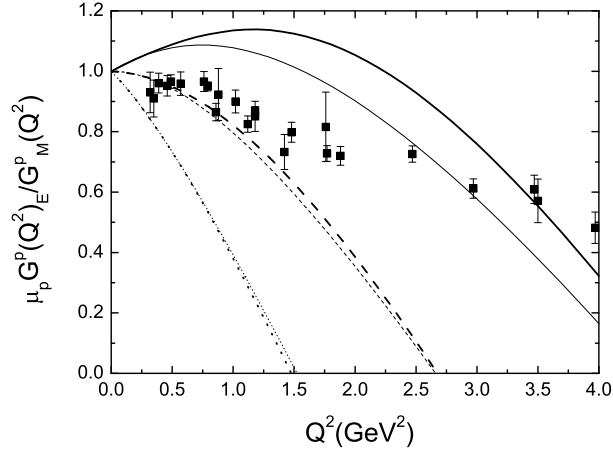}\end{center}
\caption{The ratio $\mu G^p_E/G^p_M$. The thick and thin lines are
for the Gaussian type and negative power type wave functions with
$f_V=80~MeV$ (full), $100~MeV$ (dashed), $130~MeV$ (dotted)
respectively. Experimental data are from Refs.~\cite{Jones,Gayou01,Gayou02}.}
 \label{Fig: GEpGMp}
\end{figure}

In Fig.~(\ref{Fig: GEpGMp}), we present the ratio $\mu G^p_E/G^p_M$
with different $f_V$ compared with recent data from Jefferson Lab.
With all $f_V$, the ratio will decrease in the higher $Q^2$ region
and the falloff is too fast. There are similar results for the
three-quark model in Ref~\cite{Araujo}. The difference between the
data and experiments may arise from the simplistic wave functions we
use and omission of the intrinsic form factors of quarks and
diquarks. In Fig.~(\ref{Fig: GEpGMp}) we also find that the momentum
transfer $Q^2,$ where the electric form factor of the proton crosses
zero, increases with increasing $f_V$.

\section{Summary and Discussion}

We have given a relativistic diquark model of the nucleon using the KW method.
We calculate the electromagnetic form factors of the nucleon taken as a scalar
and axialvector diquark bound state. In this work we do not consider the effect
 of scalar-vector transitions and pion or gluon exchange effects, which may
improve the numerical results. The charge and magnetic radii and
magnetic moment of the proton can be reproduced, while the magnetic
moment of the neutron is too small. The shape of the form factors
$G^p_M(Q^2)/\mu_p$ and $G^n_M(Q^2)/\mu_n$ can be reproduced also.
The zero charge form factor of neutron at $Q^2=0$ (neutron charge)
is guaranteed by the formalism we used.

There will be a zero mode in electromagnetic form factors of the
nucleon arising from the instantaneous part of the quark propagator
if we adopt $\varphi(x_2,\bm{k}_{2\perp})$ to let $h(k_2)$ have
$(1-x)^{n}$ with $n\leq0$. In our numerical calculation, we find
that the zero mode cannot be neglected. Hence the
quark-diquark-nucleon vertex functions must be chosen to remove the
zero mode to ensure the correct results in the Drell-Yan-West frame.

\Large\begin{flushleft} {\bf
Acknowledgements}\end{flushleft}\normalsize

We thank Tobias Frederico for a useful discussion. This work is supported by
the National Natural Science Foundation of China (No. 10075056 and No.
90103020), by CAS Knowledge Innovation Project No. KC2-SW-N02.

\end{document}